\documentstyle[twocolumn,aps,pra,array]{revtex}
\begin{document}
\draft
\title{Addendum to ``Quantum key distribution
without alternative measurements''}
\author{Ad\'{a}n Cabello\thanks{Electronic address:
adan@cica.es}}
\address{Departamento de F\'{\i}sica Aplicada II,
Universidad de Sevilla, 41012 Sevilla, Spain}
\date{\today}
%First version: 12 September 2000.
%This version: 8 June 2001.
%After PRA proofs
%Reply: Phys. Rev. A 63, 036302 (2001).
%Addendum: Phys. Rev. A 64, 024301 (2001).
%quant-ph/0009051v2.
\maketitle
\begin{abstract}
Recently, Zhang, Li, and Guo have proposed a particular
eavesdropping attack
[Phys. Rev. A {\bf 63}, 036301 (2001)] which
shows that my quantum key distribution protocol based on
entanglement swapping [Phys. Rev. A {\bf 61}, 052312 (2000)]
is insecure. However, security against this attack can be
attained with a simple modification.
In addition, a simpler version of the protocol using four qubits
is introduced.

\end{abstract}
%Quantum communication
%03.65.Ta}
%Foundations of quantum mechanics; measurement theory

\narrowtext
The particular eavesdropping attack proposed
by Zhang, Li, and Guo \cite{ZLG00} shows that the
protocol for quantum key distribution introduced in Ref.~\cite{Cabello00b}
is insecure.
Here I shall describe a way
to elude this eavesdropping attack.
I shall use the same notation and scenarios
as in Refs.~\cite{ZLG00,Cabello00b}.
In particular, by ``the Bell operator'' I mean one
that has eigenvectors $\left| 00 \right\rangle$, $\left| 01 \right\rangle$,
$\left| 10 \right\rangle$, $\left| 11 \right\rangle$,
as defined in \cite{ZLG00,Cabello00b}.

Consider six qubits numbered 1 to 6. Alice prepares
qubits 1 and 2 in the state
$\left| 00 \right\rangle_{12}$,
and qubits 3 and 5 in the state
$\left| 00 \right\rangle_{35}$.
Bob, prepares qubits 4 and 6 in the state
$\left| 00 \right\rangle_{46}$.
Therefore,
the initial state of the six qubits is
\begin{equation}
\left| \psi \right\rangle =
\left| 00 \right\rangle_{12} \otimes
\left| 00 \right\rangle_{35} \otimes
\left| 00 \right\rangle_{46}.
\label{inicial2}
\end{equation}
Alice sends qubit 2 out to Bob
and Bob sends qubit 6 out to Alice
using a public channel.
Then Alice randomly chooses between
the following two procedures:

(i) Alice measures the Bell operator on qubits 1 and 3.
The result of this measurement will define the key.
Then Alice measures the Bell operator on qubits 5 and 6,
and publicly announces this result.
Then Bob measures the Bell operator on qubits 2 and 4.
If Eve is not present,
the result of this measurement and
the public result would allow him to ascertain the key
(see Table I).

(ii) Alice performs the unitary transformation
\begin{equation}
S = {1 \over {\sqrt 2}}\left( {\matrix{1&1\cr
1&{-1}\cr
}} \right)
\end{equation}
on qubit 3. This transformation changes the initial
state of qubits 3 and 5, $\left| 00 \right\rangle_{35}$,
into the state
\begin{equation}
\left| ++ \right\rangle_{35} =
{1 \over \sqrt{2}}
\left( {
\left| 01 \right\rangle_{35} +
\left| 10 \right\rangle_{35}
} \right).
\label{Chip}
\end{equation}
Then Alice measures the Bell operator on qubits 1 and 3
(the result of this measurement will define the key), and
then she measures the Bell operator on qubits 5 and 6,
and publicly announces this result.
In addition, Alice announces
that she has performed the
unitary transformation $S$ on qubit 3.
In this case, Bob performs the unitary transformation $S$ on qubit 4,
and measures the Bell operator on qubits 2 and 4
(again, if Eve is not present
the result of this measurement and
the public result allow him to ascertain the key;
see Table I).

Now we will see that Eve's attack suggested in \cite{ZLG00}
does not work if Alice chooses procedure (ii). We shall
use the same numbering for the qubits as in \cite{ZLG00}.
Suppose that Alice performs the transformation $S$ on qubit 3,
and then measures
the Bell operator on qubits 1 and 3, and obtains the
result ``00.'' Then, the state of qubits 2 and 5 becomes
$\left| ++ \right\rangle_{25}$. Suppose that
when Eve measures the Bell operator on qubits 6 and 8,
she obtains
the result ``01'' (then the state of qubits 4 and 7 becomes
$\left| 01 \right\rangle_{47}$). Then Eve performs the
transformation $Z$ on qubit 2,
as suggested in \cite{ZLG00}. This transformation changes the
initial state of qubits 2 and 5, $\left| ++ \right\rangle_{25}$,
into the state
\begin{equation}
\left| -+ \right\rangle_{25} =
{1 \over \sqrt{2}}
\left( {
\left| 00 \right\rangle_{25} -
\left| 11 \right\rangle_{25}
} \right).
\label{Omep}
\end{equation}
Therefore, when Alice measures the Bell operator on qubits 2 and
5, the probability of obtaining the result ``00''
is ${1 \over 2}$, and
the probability of obtaining the result ``11'' is ${1 \over 2}$.
Suppose Alice obtains ``11.'' Then, she announces this result
and the fact that she has performed the transformation $S$; i.e.,
that she has chosen the procedure (ii). Then Bob performs the
transformation $S$ on qubit 4. This transformation
changes the initial state of
qubits 4 and 7, $\left| 01 \right\rangle_{47}$, into
$\left| -+ \right\rangle_{47}$. Therefore, when Bob
measures the Bell operator
on qubits 4 and 7, the probability of obtaining the result ``00''
is ${1 \over 2}$,
and the probability of obtaining the result ``11'' is ${1 \over 2}$.
Suppose Bob obtains ``00.'' Then Bob infers that,
if Eve is not present, the key is ``11.'' However, as we
have seen, the key is ``00.'' Therefore, Alice and Bob will detect
Eve if they compare their two bits.
In addition, once Eve knows
that Alice has chosen procedure (ii), she cannot ascertain whether the key
is ``00'' or ``11.''
Table II contains all the possible results of Alice's,
Bob's, and Eve's measurements for the
two procedures for a particular value of the key (``00'').
It shows that
Eve's attack suggested in \cite{ZLG00} works
if Alice chooses the procedure (i),
but gives only partial information about the key,
and allows
Alice and Bob to detect Eve in half of the runs
if Alice chooses the procedure (ii).

There exists an attack that allows Eve
to ascertain the key without being detected in case
Alice has chosen procedure (ii). This attack requires
a different Bell state measurement on qubits 6 and 8 and
different unitary operations on qubit 2.
However, this new attack does
not allow Eve to obtain the key without being detected
if Alice chooses procedure (i). The security of
the protocol relies on Eve necessarily deciding her attack
before knowing Alice's choice.
Assuming that
Alice randomly chooses between (i) and
(ii), and that Eve randomly chooses between
the two different attacks mentioned above,
Eve will be detected with a probability
$1- \left({3 \over 4}\right)^n$,
where $n$ is the number of pairs of bits
compared by Alice and Bob.
Thus if $N$ bits are tested, the probability of
Eve's detection is
$1- \left({3 \over 4}\right)^{N/2}$.
Therefore, the protocol is less efficient at detecting Eve than
BB84 \cite{BB84} or B92 \cite{B92}.
On the other hand, other interesting aspects of the original
protocol such as the fact that it delivers one key bit per
transmitted bit are retained in the new version.

So far, I have presented a modification of the protocol
\cite{Cabello00b} that avoids
the attack described in Ref.~\cite{ZLG00}.
My aim has been to change
the original proposal as little as possible.
A simpler protocol for quantum key distribution using
entanglement swapping is as follows:

(a) Alice prepares four qubits in the state
\begin{equation}
\left| \Psi \right\rangle =
\left| 00 \right\rangle_{12} \otimes
\left| 00 \right\rangle_{34},
\label{inicial4}
\end{equation}
and sends qubits 2 and 4 to Bob through an
insecure quantum channel.

(b) Alice chooses between the following procedures:
(I) doing nothing, (II) performing $S$ on qubit 1
(which changes $\left| 00 \right\rangle_{12}$ into
$\left| ++ \right\rangle_{12}$).

(c) Alice measures de Bell operator on qubits 1 and 3.
The result of this measurement defines the key.
Then she publicly announces the procedure she has chosen.

(d) If Alice chose (I),
then Bob would measure the Bell operator on
qubits 2 and 4; if Alice chose (II),
then Bob would perform $S$ on qubit 2
(which changes $\left| ++ \right\rangle_{12}$ into
$\left| 00 \right\rangle_{12}$), and
would then measure the Bell operator
on qubits 2 and 4.
In both cases, the result of the Bell operator
measurement would allow him to ascertain the key.

In this four-qubit version of the protocol,
for each of Alice's procedures,
Eve can ascertain the key without being detected
if she performs
the appropriate Bell state measurement on qubits 2 and 4, and
then gives them to Bob in the resulting Bell state.
However, as in the six-qubit version, such attack
would only work
for one of the procedures, and
it could be detected in half of the runs
of the other procedure.

Both the revised protocol and the new one are secure against a
specific attack. It would be desirable to obtain rigorous
proofs of security analogous to the proofs of
security of the BB84 \cite{LC99,Mayers98,BBBMR00,SP00}.
Indeed, it may not be unfeasible to obtain such
proofs as a derivation of the existing ones, since both the revised
protocol and the new one have greater similarities with the
BB84.\\

I wish to thank
Y.-S. Zhang, C.-F. Li, and G.-C. Guo
for pointing out the insecurity of
the protocol \cite{Cabello00b},
for checking the modification
presented here, and for
suggesting a way to simplify it.

\newpage
\onecolumn

\begin{table}
\begin{center}
\begin{tabular}{lcccc}
\hline
\hline
\multicolumn{3}{c} {Alice} &
\multicolumn{2}{c} {Bob} \\ \hline
Procedure$\;$ & $\;$Key$\;$ & $\;$Public$\;$ &
$\;$Secret$\;$ & $\;$Inferred key$\;$
\\ \hline
(i) & 00 & 00 & 00 & 00 \\
(i) & 00 & 01 & 01 & 00 \\
(i) & 00 & 10 & 10 & 00 \\
(i) & 00 & 11 & 11 & 00 \\
(ii) & 00 & 00 & 00 & 00 \\
(ii) & 00 & 10 & 01 & 00 \\
(ii) & 00 & 01 & 10 & 00 \\
(ii) & 00 & 11 & 11 & 00 \\
\hline
\hline
\end{tabular}
\end{center}
\end{table}
%\vspace{0.2cm}
TABLE I. {\small The columns represent,
from left to right,
the strategy chosen by Alice,
the result of her secret measurement
on qubits 1 and 3 (the key),
the result of the public measurement
on qubits 5 and 6,
the result of Bob's secret measurement on qubits 2 and 4,
the result that Bob thinks is the key,
in all the cases
in which the key is ``00,'' and assuming that
the initial state is given by Eq.~(\ref{inicial2}).}

\bigskip

\begin{table}
\begin{center}
\begin{tabular}{lccccccc}
\hline
\hline
\multicolumn{3}{c} {Alice} &
\multicolumn{2}{c} {Bob} &
\multicolumn{3}{c} {Eve} \\ \hline
Procedure$\;$ & $\;$Key$\;$ & $\;$Public$\;$ &
$\;$Secret$\;$ & $\;$Inferred key$\;$ &
$\;$Secret$\;$ & $\;$Transformation$\;$ & $\;$Inferred key$\;$
\\ \hline
(i) & 00 & 00 & 00 & 00 & 00 & $I$ & 00 \\
(i) & 00 & 01 & 01 & 00 & 01 & $Z$ & 00 \\
(i) & 00 & 10 & 10 & 00 & 10 & $X$ & 00 \\
(i) & 00 & 11 & 11 & 00 & 11 & $Y$ & 00 \\
(ii) & 00 & 00 & 00 & 00 & 01 & $Z$ & 00 or 11 \\
(ii) & 00 & 11 & 00 & 11 & 01 & $Z$ & 00 or 11 \\
(ii) & 00 & 00 & 00 & 00 & 10 & $X$ & 00 or 11 \\
(ii) & 00 & 11 & 00 & 11 & 10 & $X$ & 00 or 11 \\
(ii) & 00 & 01 & 01 & 11 & 00 & $I$ & 00 or 11 \\
(ii) & 00 & 10 & 01 & 00 & 00 & $I$ & 00 or 11 \\
(ii) & 00 & 01 & 01 & 11 & 11 & $Y$ & 00 or 11 \\
(ii) & 00 & 10 & 01 & 00 & 11 & $Y$ & 00 or 11 \\
(ii) & 00 & 01 & 10 & 00 & 00 & $I$ & 00 or 11 \\
(ii) & 00 & 10 & 10 & 11 & 00 & $I$ & 00 or 11 \\
(ii) & 00 & 01 & 10 & 00 & 11 & $Y$ & 00 or 11 \\
(ii) & 00 & 10 & 10 & 11 & 11 & $Y$ & 00 or 11 \\
(ii) & 00 & 00 & 11 & 11 & 01 & $Z$ & 00 or 11 \\
(ii) & 00 & 11 & 11 & 00 & 01 & $Z$ & 00 or 11 \\
(ii) & 00 & 00 & 11 & 11 & 10 & $X$ & 00 or 11 \\
(ii) & 00 & 11 & 11 & 00 & 10 & $X$ & 00 or 11 \\
\hline
\hline
\end{tabular}
\end{center}
\end{table}
%\vspace{0.2cm}
TABLE II. {\small The columns represent,
from left to right,
the strategy chosen by Alice,
the result of her secret measurement
on qubits 1 and 3 (the key),
the result of the public measurement
on qubits 5 and 6
(5 and 2 in the notation of Ref.~\cite{ZLG00}),
the result of Bob's secret measurement on qubits 2 and 4
(7 and 4 in the notation of Ref.~\cite{ZLG00}),
the result that Bob thinks is the key,
the result of Eve's measurement
on qubits 6 and 8 (in the notation of Ref.~\cite{ZLG00}),
the corresponding unitary transformation on qubit 2
(as described in Ref.~\cite{ZLG00}),
the result that Eve thinks is the key,
in all the cases
in which the key is ``00,'' and assuming that
the initial state of Alice's and Bob's qubits
is given by Eq.~(\ref{inicial2}),
and that the initial state of Eve's qubits
is $\left| 00 \right\rangle_{78}$.}

\end{document}